\documentclass{JHEP}
\usepackage{epsfig}

\title{Multivalued Entropy of \\
Supersymmetric Black Holes } 

\author{Renata Kallosh\thanks{On leave of
absence from Stanford University until 1 September 2000}\\ 
    Theory Division, CERN CH 1211 Geneve 23, Switzerland\\
    E-mail: \email{Renata.Kallosh@cern.ch}
} \received{\today} 

 \preprint{CERN-TH/99-378\\7 December 1999\\
  hep-th/9912053}

\abstract{ The supersymmetric flow equations describing the flow of 
moduli from infinity to the black hole horizon, and vice versa, are 
derived in the five-dimensional theories where the moduli space of the 
very special geometry has disjoint branches. The multiple solutions are 
derived from the `off the horizon' attractor equation.  
 Within each branch, the black hole
entropy, as usual, depends only on the near horizon attractor values of 
moduli, i.e. the entropy  depends on the charges and on coefficients 
 of the cubic polynomial. It
 does not depend on the values of the moduli fields at 
infinity. However, the entropy, as well as the near horizon values of the 
moduli fields, are shown to depend on the choice of the branch specified 
by the choice of the set of moduli at infinity. We present examples of 
BPS black hole solutions with the same  $Q_I$ and $C_{IJK}$, whose 
 entropies differ significantly.} 


\begin{document}

1. During the last decade supersymmetric black holes played an important 
role in setting up  the issues of the fundamental theory, including 
gravity. For example, an important property of the supersymmetric black 
holes is the manifest symmetry under $U$-duality transformations. The 
explanation of this fact from the point of view of type IIA string theory 
requires first to promote it to an 11-dimensional theory and afterwards 
to compactify it on a torus. Not much is known about 11-dimensional M 
theory, but the existence of  $U$-duality invariant BPS states, like 
supersymmetric black holes, helps to  explore the non-perturbative 
string/M-theory. 

The purpose of this note is to present some new, previously unexplored 
features of supersymmetric black hole entropy: 
 its non-uniqueness in 
theories with disjoint branches of the moduli space.  The existence of 
multiple critical points relevant to BPS black holes in the theories with 
the same electric charges but disjoint branches of moduli space was 
already established before \cite{KLS}. 

The issue of the non-uniqueness of the supersymmetric black hole entropy 
 was raised in 
\cite{moore} in the context of black holes of  Calabi-Yau spaces. Here we 
consider arbitrary d=5, N=2 supergravity theory \cite{GST}, the moduli 
space is not symmetric, in general. 
 We put no restrictions on  
Chern-Simons couplings $C_{IJK}$: the supersymmetry is valid for 
 all these generic theories. In some cases they may be interpreted as 
 Calabi-Yau intersection numbers. 
  The uniqueness of the entropy 
in the theories where the moduli space has only one branch with the 
positive metric to large extent follows from the fact, established in 
\cite{GFK}, that the entropy is a minimum of the 
 BPS mass.\footnote{
Recently   a detailed derivation of the uniqueness of the BPS black hole 
entropy was performed in  \cite{WZ}. The claim was also made that the 
entropy is  always unique. The argument, however, is based on an  
assumption that the  moduli space consists of a single branch.} 
 The existence of disjoint branches of 
 the moduli space of five-dimensional supergravity \cite{GST}
  was first pointed out in \cite{GST2}.

 In the context of gauged supergravity   the  disjoint  branches 
of moduli space were studied in \cite{BC,KLS}. An important feature of 
these branches is that not only the metric of scalar fields \cite{BC} but 
also the metric of vector fields is positive-definite \cite{KLS}, which 
means that these branches are quite legitimate. In particular, we have 
found different $AdS_5$ branches with equal values of the cosmological 
constant, which is one of the necessary conditions for realization of a 
supersymmetric generalization of the one-brane Randall-Sundrum scenario 
\cite{RS}. However, we were able to show (see also \cite{BC}) that all  
interpolating domain wall solutions  (not only BPS states) in a broad 
class of supersymmetric models studied in \cite{KLS} are not of  the 
Randall-Sundrum type.

In this paper we will present the complete black hole solutions, i.e. 
define the moduli and the metric and the vector fields of the black holes 
everywhere, not only at the critical point near the horizon. There will 
be at least two solutions depending on the same harmonic functions, one 
with the central charge  $Z$ positive and the other with the central 
charge $Z$ negative.  This will explain the multivalued nature of the 
black hole entropy.

 We study  the five dimensional N=2 ungauged  supergravity interacting 
with $n$  abelian vector multiplets \cite{GST,VdW}. The supersymmetric 
black hole solutions of this theory in the context of the very special 
geometry were studied in \cite{CFGK,CKRRSW}. A general ansatz for the 
black hole solution with positive central charge $Z$ was proposed in 
\cite{Sabra}. The ansatz is given in terms of $n+1$ harmonic functions, 
$K_I= k_I + Q_I/r^2$. It provides an explicit  black hole solution only 
in cases when the solution of the stabilization-type equations 
$C_{IJK}Y^J Y^K= K_I$  is available.   Stabilization equations in general 
are known to define the values of the moduli near the black hole horizon 
\cite{FKS,FK}. The specific form of stabilization equations in N=2 d=5 
supergravity interacting with vector multiplets, $C_{IJK}Y^J Y^K= Q_I$, 
was found in \cite{CKRRSW}. It gives the values of the fixed scalars near 
the horizon. An analogous equation is  also a part of the 
 stabilization equations in  d=4 N=2 theory  with a cubic  prepotential
 \cite{Marina} relevant to Calabi-Yau black holes.

We will present below an ansatz for the black hole solutions with both 
positive and negative central charges $Z=X^I(\phi) Q_I$. The surprising 
feature of it is that sometimes both solutions with positive and negative 
central charges (graviphoton charges)  occur for the same choice of 
harmonic functions and in particular for the same individual vector 
fields charges $Q_I$.

The reason why it is surprising to have a central charge, i.e. the 
graviphoton charge, both positive and negative for the same charges of 
the vector fields $Q_I$ is the following. The graviphoton charge is given 
by the moduli dependent combination  $Z=X^I(\phi) Q_I$ of the individual 
vector fields charges $Q_I$. If there are no vector multiplets and $I=0$, 
the graviphoton charge is equal to the usual charge $Z=Q_0$. The central 
charge is positive for positive $Q_0$ and negative for negative $Q_0$. In 
the first case we have $M=Z$, in the second case $M=-Z$. In presence of 
moduli, as shown in \cite{KLS} it is possible to have both values of $Z$ 
without changing the sign of $Q_I$. After having observed this unusual 
situation near the critical point we would like to show the full solution 
with such properties.

\

2. The very special geometry of five-dimensional supergravity emerges 
because the independent moduli $\phi^i,\; i=1, \dots ,n$ are coordinates 
describing some cubic hypersurface 
\begin{equation}
{\cal V} = {1\over 6} C_{IJK} X^I X^J X^K =1\ , \qquad I=0, 1, \dots  , 
n. \label{hypersurface} 
\end{equation}
The five-dimensional bosonic  N=2 Lagrangian is: 
\begin{equation}
e^{-1} {\cal L} = -{1\over 2} R - {1\over 4} G_{IJ} F_{\mu\nu} {}^I 
F^{\mu\nu J}-{1\over 2} g_{ij} \partial_{\mu} \phi^i \partial^\mu \phi^j 
+{e^{-1}\over 48} \epsilon^{\mu\nu\rho\sigma\lambda} C_{IJK} 
F_{\mu\nu}^IF_{\rho\sigma}^JA_\lambda^k \ . 
\end{equation}
The  gauge coupling metric $G_{IJ}$ and the moduli space metric $ g_{ij}$ 
are 
\begin{equation}
G_{IJ}(\phi) = -{1\over 2}{\partial\over \partial X^I} 
{\partial\over\partial X^J}(\ln {\cal V})|_{{\cal V} =1}  \ , \qquad 
g_{ij}(\phi)=  G_{IJ} 
\partial_{i}X^I\partial_{j}X^J|_{{\cal V} =1} \ .
\label{metrics} 
\end{equation}

We are looking for the electrically charges black holes with the 
 metric 
\begin{equation}
  ds^2 =-e^{-4 U} dt^2 +e^{2 U} (d\vec{x})^2 \ .
\label{metric} 
\end{equation}
The Chern-Simons term does not contribute to electric configuration where 
only $F_{0r}^I$ are not vanishing.  We  introduce  a function ${\cal Z} 
\equiv   X^I G_{IJ} F^J_{or}$ and a function ${\cal Z }_i \equiv 
\partial_i X^I G_{IJ} F^J_{or}$ and consider configurations for which
 ${\cal Z }_i = \partial_i{\cal Z }$.
The action can be rewritten in a nice BPS form. 
\begin{eqnarray}
&&E = -\int dr_{-\infty }^{+\infty} {\cal L}= \nonumber\\ &&\int 
_{-\infty }^{+\infty} dr\left [{3\over 2} \left ({\partial U\over 
\partial r} \pm {1\over 3} e^{2U} {\cal Z}\right)^2 + {1\over 2}
 \left ({f_{ai} \partial \phi^i\over \partial r} \pm {1\over 2}e^{2U} f_a^ i
 {\cal Z}_{i} \right)^2
 \mp {1\over 2} {\partial\over \partial r} \left( e^{2U} {\cal Z}
 \right)\right].
 \label{BPS}
\end{eqnarray}
Here we used the moduli space vielbein $f_i^a$ where $ f_i^a 
f_j^b\delta_{ab}  =g_{ij} $. The analogous form of the action in case of 
4-dimensional black holes was found in \cite{GFK}. One can use either 
supersymmetry or just  very special geometry to derive eq. (\ref{BPS}). 

The first order  flow equations which define the evolution  of  the 
metric function $U(r)$  and of the moduli $\phi^i(r)$ follow from this 
action 
\begin{eqnarray}
 {\partial U\over
\partial r} \pm {1\over 3} e^{2U} {\cal Z}=0 \ , \qquad
 {g_{ij} \partial \phi^j\over \partial r} \pm {1\over 2}e^{2U}
 {\cal Z}_{i}=0 \ .
 \label{flow}
\end{eqnarray}
Solutions of these equations  with the boundary conditions for which the 
surface term vanishes saturate the BPS bound. The solutions are given by 
\begin{eqnarray}
ds^2 &=&-e^{-4 U} dt^2 +e^{2 U} (d\vec{x})^2 \ , \\ 
 G_{IJ} F_{0m}^I &=& {1\over4}e^{-4U}\partial_m K_J \ ,\\
  \label{ansatz}
  e^{2U} &=& \pm {1\over 6} X^I K_I \ .
  \label{positive} 
\end{eqnarray}
Here $K_I$ is a harmonic function, 
\begin{equation}    K_I= k_I+{Q_I\over r^2}\ ,
\label{harmonic} 
\end{equation}
and the moduli $X^I(\phi)$ are real and have to satisfy the `off the 
horizon'  attractor equation 
\begin{eqnarray}
\pm e^{2U}  C_{IJK} X^J X^K = K_I \ . \label{off} 
\end{eqnarray}

The ansatz with the upper sign is the one found by Sabra \cite{Sabra}. 
The appearance of the second solution with the minus sign for the same 
choice of the harmonic function is new since we are not changing $K_I$ to 
$-K_I$. It is clear from eq. (\ref{positive}) that $e^{2U}$ must be 
everywhere positive since it is a component of the space-time metric. 
Still the combination $X^I K_I$ can  be either positive or negative: only 
in such cases our ansatz  gives a consistent solution. Both cases may 
exist as we will see soon. 

Near the black hole horizon our `off the horizon' attractor equation 
(\ref{off}) reduces to the near horizon  attractor equation 
\begin{equation}
 { Z_{hor}\over 6}\, C_{IJK}
X^J X^K =  Q_I \ . \label{nearhorizon} 
\end{equation}
To see this one has to use the fact that at $r\rightarrow 0$ the metric 
tends to $e^{2U}\rightarrow {|Z|_{hor}\over 6 r^2}$. 

If we absorb the central charge into the redefinition of the moduli near 
the horizon, $\bar X^I = \sqrt{ Z_{hor}/6} X^I$, we may bring the near 
horizon stabilization equation  to the form 
\begin{equation}
  C_{IJK}\bar X^I_{hor}\bar  X^J_{hor} = Q_I \label{nearhor} 
\end{equation}

In this form it was studied extensively in \cite{CKRRSW} and, for 
 the multivalued central charge  $Z$, in \cite{KLS}.

The multivalued nature of the black hole entropy was discussed in 
\cite{KLS} on the basis of the near horizon attractor equation 
(\ref{nearhor}). At least 2 
 different 
solutions may exist: one with positive $Z$ and another one with negative 
$Z$. The fields $\bar X^I$ may be real and imaginary, in what follows we 
will denote them by $Y^I_{re}$ and $ Y^I_{im}$, respectively. 
\begin{equation}
  \sqrt{+ 
 |Z_{hor})|/6}\; X^I_{hor} = Y^I_{re}(r=0)\ , \qquad \sqrt{ - 
 |Z_{hor}|/6}\; X^I_{hor} = Y^I_{im}(r=0)\ .
\label{redef} 
\end{equation}
In the first case the redefined field is real, in the second case 
 it is imaginary. However, the moduli $X^I$ in both cases is real. The 
 near horizon
 equations  take the form
\begin{eqnarray}
  C_{IJK} Y^J_{re} Y^K_{re} = Q_I \ ,  \qquad  C_{IJK} Y^J_{im} Y^K_{im} =
   Q_I   \ .
   \label{near}
\end{eqnarray}

 Even more 
 solutions may exist 
 in many-moduli case, particularly if the stabilization equations 
 (\ref{nearhor}
   have 
 more solutions for which the metric of the moduli space and the one for 
 the vector space are positive-definite.  
 We will see now how all of this generalizes for the full black hole 
 solution. First we redefine the fields in eq. (\ref{off}). We introduce
 $\sqrt {\pm e^{2U}} X^I= \bar X^I (r)$. In terms of these variables the 
 `off the horizon' attractor equation takes the form: 
\begin{equation}
  C_{IJK}\bar X^J(r)\bar  X^K(r) = K_I(r) \ . \label{nearhor2} 
\end{equation}

 Introduce 
\begin{equation}
\sqrt {+ e^{2U}} X^I= Y^I_{re}(r) \ , \qquad  \sqrt {- e^{2U}} X^I= 
Y^I_{im}(r) \ . \label{redefini} 
\end{equation}
The off the horizon attractor equations take the form 
 \begin{eqnarray}
  C_{IJK} Y^J_{re}(r) Y^K_{re}(r) = K_I(r) \ ,  \qquad  C_{IJK} Y^J_{im}(r) Y^K_{im}(r) =
   K_I(r)  \ . 
   \label{off1}
\end{eqnarray}

Here again we will be looking for the solutions of the stabilization 
equations with both real and imaginary $Y^I$ which correspond to real 
$X^I$. 
 Note that after the field redefinition the  off the 
 horizon equations (\ref{off1}) look exactly as the near horizon equations  (\ref{near})
but now  the moduli $Y(r)$ are defined by its solutions via $K_I(r)$ 
everywhere at 
 all values of $r$ and not only at $r=0$. Still the solutions of the
  attractor equations for 
 $Y^I_{re}(r)$ and $Y_{im}^I(r)$ in terms of the harmonic functions
 $K_I(r)$ are precisely the same as in the near horizon 
 cases the solutions for  $Y^I_{re}(r=0)$ and $Y_{im}^I(r=0)$ 
 in terms of the constants $Q_I$.
This  means that in all cases when these equations near horizon 
 were  solved (or will be solved eventually), and multiple solutions are 
 available 
  near the  horizon,  
 the multivalued black hole solution are obtained simply by replacing the 
 constants $Q_I$ 
 with the harmonic functions  $K_I(r)$. In particular, using examples 
of solutions with both real and imaginary 
 values of $Y^I(r=0)$ given in 
  \cite{KLS},  we will find below the full black hole solutions with 
 both real and imaginary $Y^I(r)$ and two values of real $X^I(r)$.

Thus,  to find the black hole solutions in these theories one 
should   explicitly solve the algebraic attractor equations. The metric will be 
obtained straightforwardly from such solutions. 

There is an interesting property of our attractor equations at 
$r\rightarrow \infty$. Since the parameters of the solutions are given by 
harmonic functions $H_I$ the values of moduli at $r\rightarrow \infty$ 
are not free: rather we have to solve the stabilization equation at  
$r\rightarrow \infty$ to find the values of the moduli there. 
 \begin{eqnarray}
  C_{IJK}( Y^J_{re} Y^K_{re})_{r\rightarrow \infty}  = k_I \ , 
   \qquad  C_{IJK}( Y^J_{im} Y^K_{im})_{r\rightarrow \infty} = k_I \ .
   \label{offinf}
\end{eqnarray}
 There is a restriction on the choice of $k_I$ for a 
given $C_{IJK}$ such that for both solutions $e^{2U}\rightarrow 1$ at 
$r\rightarrow \infty$: 
\begin{equation}
  e^{2U}_{r\rightarrow \infty} = \pm {1\over 
6}( X^I_{r\rightarrow \infty})\;  k_I \label{infty} =1 \ . 
\end{equation}

The double extreme black holes are the ones with constant moduli. To find 
such  we may choose $k_I=\alpha Q_I$ and all harmonic functions will have 
the factorizable dependence on $r$ of the form $K_I= k_I( 1+\alpha 
/r^2)$. The moduli fields if chosen e. g. as the ratio of $X^I/X^0$ will 
be $r$-independent for such solutions and the metric will be $e^{2U}=( 
1+\alpha /r^2)$. We will find examples of such double-extreme black holes 
which live in the two different branches of the moduli space. The values 
of moduli will be different in two branches, but the metric and therefore 
the entropy will be the same.

For the  black hole solutions with non-constant moduli, the value of the 
metric at infinity defined by the choice of $h_I$ will be the same.  
However at the horizon the metric in two different branches of the moduli 
space and, consequently,  the black hole entropy, can be significantly 
different.

\

3. The theory with one independent moduli is relatively easy to 
understand. It may give some insights into the properties of the black 
holes in more general and more interesting cases with many moduli, in 
particular, related to Calabi-Yau spaces. The positive definiteness of 
the moduli space metric and the gauge couplings has been analyzed in 
one-moduli case in \cite{KLS}. In more general situations this will be a 
more complicated problem.

 Consider a simple case of  $I=1,2$ and generic $C_{IJK}$ and
  $H_I= {1\over 2}K_I $. According to the discussions above, we 
  have to take our solutions of the near 
  horizon attractor equations in \cite{KLS} and replace the charges there 
  by the harmonic functions $ K_I(r)$ to get the full metric. It 
  gives for the two cases of $Y^I_{re}$ and $Y_{im}^I$ the solutions for 
  the metric and for 
  the moduli $\phi\equiv {X^2\over X^1}$:
  \begin{eqnarray}
\mp \left(e^{6U}(r)\right)_{P/M} & =& -{K _2 (d K _1^2+bK_2^2-2cK_1K _2) 
\over 36 M}+\nonumber\\ 
 &+& {\cal D} 
\left[{F(K)L +2E(K) M  \pm  \sqrt{4 M^2 {\cal D}(K) }\over 36 M( L^2- 
4MN)} 
 \right ], 
\label{metricsolution} \end{eqnarray} 

\begin{equation}
 \left( \phi(r)\right)_{P/M} = {-E(K) \pm \sqrt{{\cal  D}(K)} \over F(K)} 
 \ .
\label{moduli} 
\end{equation}
Here\footnote{We assume that $M\neq 0$ and $L^2 \neq 4MN$.} 
\begin{equation}
C_{111}=a\ ,\quad C_{112}=b\ ,\quad C_{122}= c \ ,\quad C_{222}= d \ , 
\end{equation}
\begin{equation}
M \equiv c^2-bd\ , \qquad N \equiv b^2-ac \ , \qquad  L \equiv ad-bc \ , 
\end{equation}
\begin{equation}
{\cal D}(K) \equiv (MK _1^2+ N K _2^2+ L K_1 K _2) \ , 
\end{equation}
\begin{equation}
E (K)\equiv cK_1 -bK_2\ ,    \qquad F(K) \equiv d K_1 -c K_2 \ . 
\end{equation}
The intermediate expressions for $Y^1_{re}, Y^2_{re}$ and $Y^1_{im}, 
Y^2_{im}$ can be found using \cite{KLS} and performing the replacement of 
charges by harmonic functions.   We derived the solutions above using the 
following definitions: 
\begin{equation}
  \phi_P= \left( {X^2\over X^1}\right)_P={ Y^2_{im}\over Y^1_{im}} \ , 
  \qquad \phi_M =  \left({X^2\over X^1}\right)_{M}={ Y^2_{re}\over 
  Y^1_{re}} \ ,
\label{phi} 
\end{equation}
and 
\begin{equation}
 \left(e^{6U}(r)\right)_{P}= - \left({1\over 6} Y^I_{im} K_I \right)^2 \ 
 ,  \qquad \left(e^{6U}(r)\right)_{M}=  \left({1\over 6} Y^I_{re} K_I 
 \right)^2 .
 \end{equation}
\label{U} The moduli space metric is 
\begin{equation}
g_{\phi\phi}= {3[N- L\phi + M\phi^2]\over [a+3b\phi + 3c\phi^2 + d 
\phi^3]^2} \ . \label{modulimetr} 
\end{equation}
We assume that 
$
L^2- 4MN<0$ and  $ M>0$, $ N>0$  so that the metric of the moduli space 
is  positive. The metric, however, may have singularities which 
 show that the moduli space defined by the constrained surface ${\cal 
 V}=1$ has 
disjoint branches: the multivalued black hole solutions exist in each 
separate branch of the moduli space. 
 The study of the  vector space metric (gauge couplings) for 
the black holes defined above is performed in analogy with the near 
horizon case in \cite{KLS} and we find again that it is positive. 

The fixed values of moduli at infinity and at the horizon are  
\begin{equation}
 \left( \phi(r)\right)_{P/M}^{\infty} =
  {-E(k) \pm \sqrt{{\cal  D}(k) }\over F(k)}\ , \qquad 
   \left( \phi(r)\right)_{P/M}^{hor} =
  {-E(Q) \pm \sqrt{{\cal  D}(Q) }\over F(Q)} \ .
\label{mod} 
\end{equation}
For arbitrary choice of harmonic functions the values of $X^I K_I$ for 
two solutions (with $Y_{re}$ and $Y_{im}$) are different. But at 
$r\rightarrow \infty$ the metric has to be the same, 
$e^{2U}_{r\rightarrow \infty}=1$. Fortunately, solutions of such type 
have been found in \cite{KLS}. One has to require, therefore, that  by 
solving the attractor equations at infinity (\ref{offinf}) one gets the 
same value of $X^IK_I$ in case of $Y_{re}$ and $Y_{im}$. The equation for 
our parameters specifying this case is: 
\begin{eqnarray}
{\cal A} \equiv -k _2 (d k _1^2+bk_2^2-2ck_1h _2) +  {\cal D} (k) 
\left[{F(k)L +2E(k) M \over ( L^2- 4MN)} \right]=0 \ . 
 \label{cosm} \end{eqnarray}
 In \cite{KLS} we have found several families of such solutions of the 
 stabilization equations. We will use these examples to satisfy the 
 condition $ \left(e^{2U}\right)_{P/M}  =1$ at $r\rightarrow \infty $.

\

4. To show that the analytic solutions for the multivalued black holes 
have non-trivial examples and to understand the properties of such 
solutions, we plot some of them for a particular choice of $C_{IJK}$. 
 As the purpose of this paper is to promote the multiple critical 
points of our attractor systems to complete black hole solutions, we will 
use the same one-moduli theory with $a=0,~ b=1/3, ~ c=4/3, ~ d=1 $ which 
we used in \cite{KLS} and where we checked that  the physical conditions 
of the positivity of the moduli space metric and of the gauge coupling 
matrix are satisfied.  In this case the moduli space metric is singular 
 at $\phi=0, \phi \sim -0.27, \phi\sim  -3.73$.  It is 
positive everywhere but discontinuous. Therefore the expectation is that  
if the moduli at infinity starts in one of the branches, it will flow to 
the black hole horizon remaining in the same stripe of the moduli space, 
so that at all $r$ the black hole moduli $\phi(r)$ is inside of a given 
branch where it started and the metric $g_{tt}$ is smooth all the way 
from infinity to the horizon. This indeed is a property of our solutions. 
We plot some examples of multivalued black hole solutions and their 
entropy. In all our examples we take $k_1=2$, $k_2=4$. These provide the 
correct asymptotic behavior of the metric $e^{2U}\rightarrow 1$ and keep 
the initial values of moduli in two disconnected branches of the moduli 
space: the first one in the branch $0<\phi<+ \infty$ and the second one 
in the branch $-0.27<\phi<0$. In all examples we will plot the moduli 
$\phi(r)$ and the space-time metric.  We also plot the metric $g_{tt}(r) 
= e^{-4U(r)}$; this function tends to $0$ near the horizon at $r=0$ and 
tends to $1$ at $r\rightarrow \infty$. To find the entropy we plot for 
each example the value  of $r^3 e^{3U(r)} = |Z_{hor}/6|^{3/2}(r)= |\tilde 
Z_{hor}|^{3/2}(r) $ near the horizon. The black hole  entropy  is 
proportional to $|\tilde Z_{hor}|^{3/2}(0)$. 

It is  useful to start with the double extreme black holes: 
 for
 them the moduli does not change, so it definitely stays in the same branch
  where it started. 
 We make here the simplest  choice of harmonic functions $K_I= 
 k_I(1+1/r^2)$ with $k_1=2$, $k_2=4$.
 The plot of moduli $\phi$ in Figure 1   shows
  that one of them remains equal to $+1$ everywhere and the other one 
  remains equal to $-0.2$.
 In  Figure 1  we also show the metric: we find that  $g_{tt}$,
   is the same for both solutions, as follows from analytic expression
    for the
    double extreme black holes. The entropy  is also the same for these two solutions.
 
 \FIGURE{\epsfig{file=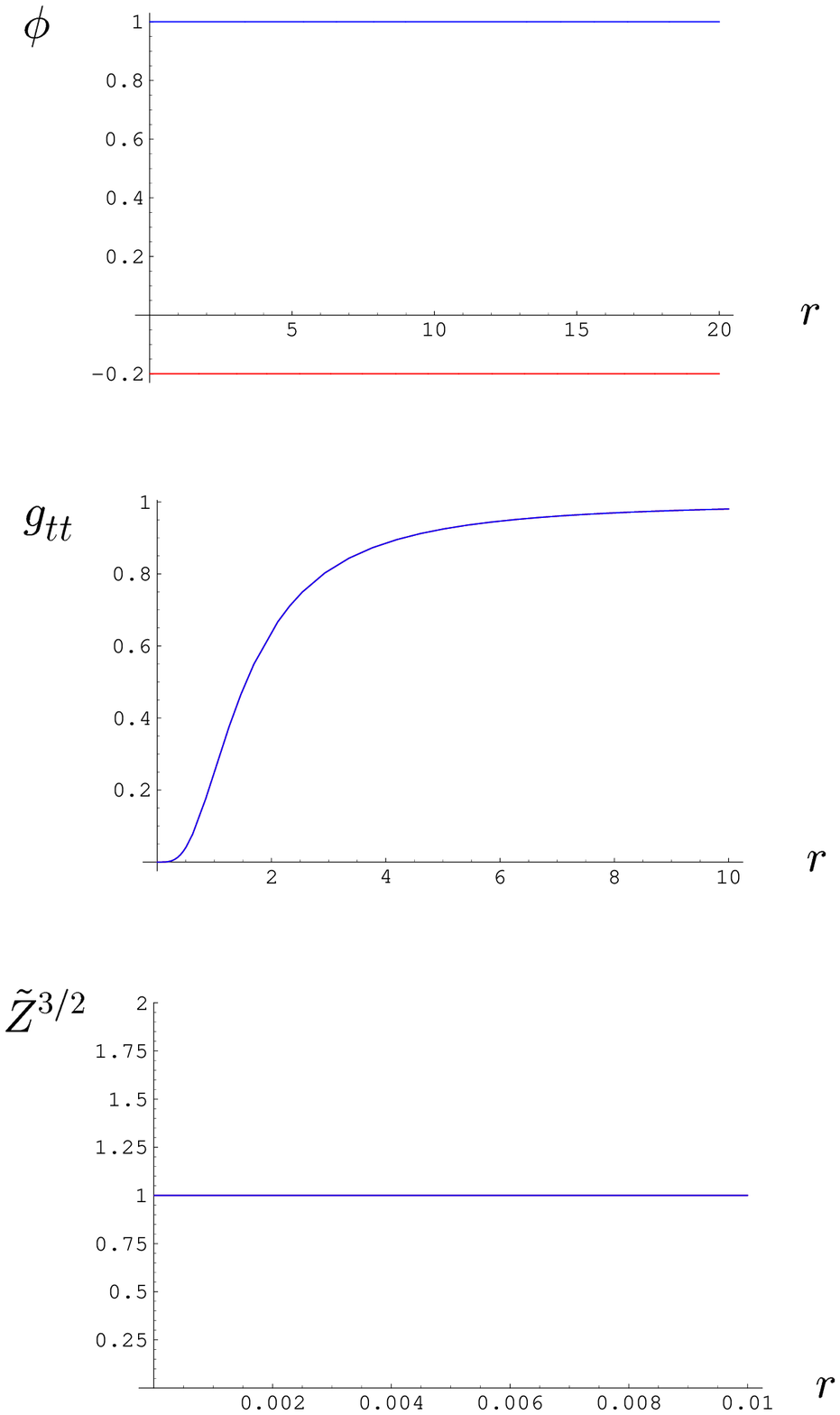,width=12cm}
        \caption{An example of a  multivalued double extreme black hole
         with
        everywhere constant moduli, which are different in two branches of 
        the moduli space.
        The metric (and therefore the entropy) for both choices of moduli are the same. 
        }
    \label{myfigure}}

The second example which we plot in Figure 2 is for a choice $ k_1 = 2,~  
k_2 = 4,~Q_1 = 0.49,~ 
 Q_2 = 1.8275$. This particular choice of charges was taken simply to 
 show that one can find a solution with the entropies near the horizon of 
 each black hole different $10^3$ times. In Figure 2 we plot the two flows of moduli, 
 each in its branch of the moduli space. To show 
 that both solutions are nice and smooth, we plot 
  $g_{tt}(r))_{P/M}=(e^{-4U(r)})_{P/M}$ on two different scales: 
  one shows that both metrics 
at infinity approach $1$, the second is closer to the horizon. 
We also plot the entropy. We have to make two different plots, since the first 
one differs from the 
second one $10^3$ times. We have plotted many other examples with simple 
 values of charges, like 1, 3, 5 etc. In such cases we found the 
entropies in two branches differ moderately, like  2 or 10 times.

  \FIGURE{\epsfig{file=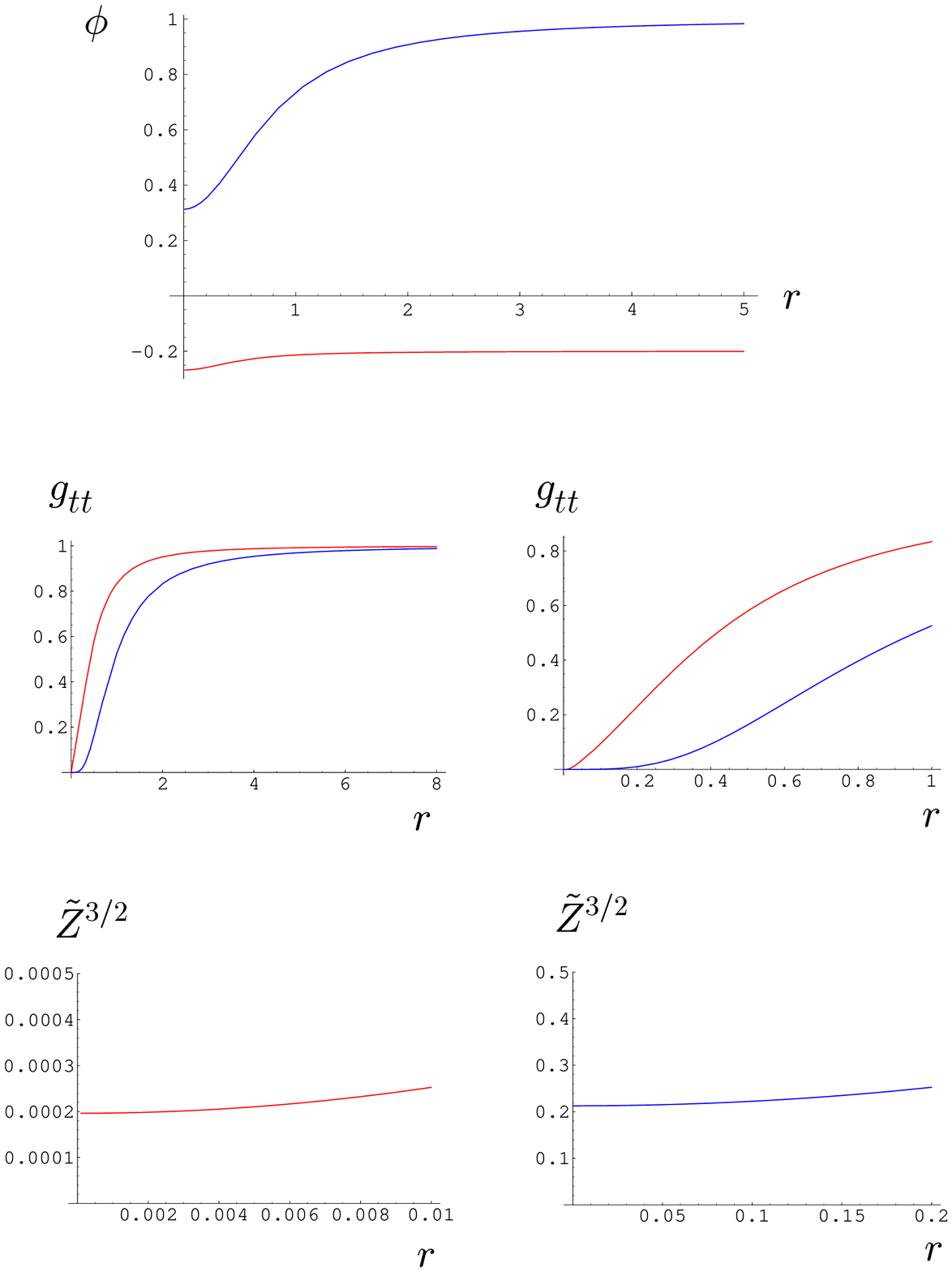,width=14cm}
        \caption{An example of a black hole solutions with two attractors, one for  
       each branch of the moduli space. The metric and the entropy differ strongly 
       near the horizon. The entropy of these black holes differs $10^3$ times. }
    \label{myfigurecr}}

Thus we have studied supersymmetric black holes in 
 d=5, N=2 supergravity in cases when the moduli space may have disjoint 
branches with everywhere positive metric. We have found  black hole 
solutions in each branch which are different despite the black hole 
charges and the cubic surface defining a supergravity theory are the 
same. The new supersymmetric black hole entropy formula is 
\begin{equation}
  S = S(Q_I, C_{IJK},N_l )\ ,
\label{entropy} 
\end{equation}
where the dependence on $N_l$ with $l=1\dots , k$ indicates the 
dependence on the branch of the moduli space in case there are $k$ 
branches. It would be interesting to find the multivalued black hole 
solutions for some parameters which may appear within a context of 
string/M-theory.

It remains to find out whether this observation of the unusual properties 
of supergravity black holes can be used to probe the properties of the 
fundamental theory.

I am grateful to A. Chamseddine, S. Ferrara, A. Linde, G. Moore, M. 
Shmakova,  and J. Rahmfeld and especially to G. Gibbons and A. Van 
Proeyen for valuable discussions. I am grateful for the hospitality to 
the Institute of Theoretical Physics in Leuven  where this work was 
finished. 
 This work 
was supported in part by NSF grant PHY-9870115.

\end{document}